\documentclass[12pt]{article}
\pdfoutput=1
\usepackage{xspace}
\usepackage{hyperref}
\usepackage{amsmath}
\usepackage{amssymb}
\usepackage{color}
\usepackage{array}
\usepackage[absolute,overlay]{textpos}
\usepackage{graphicx}
\usepackage{multirow}
\usepackage{longtable}
\usepackage{pstricks}
\usepackage{pdflscape}
\usepackage{slashed}
\usepackage{pifont}
\usepackage{ulem}

\newcommand\pubnumber{IPPP/19/4}
\newcommand\pubdate{}

\newcommand{\institute}{%
  Theoretical Physics Department, CERN, CH-1211 Geneva 23, Switzerland\\[1mm]
  \textup{and}\\[1mm]
  Institute for Particle Physics Phenomenology, Department of Physics, Durham University, Durham, DH1 3LE, UK
}

\newcommand{\Title}[1]{\begin{center} {\Large #1 } \end{center}}
\newcommand{\Author}[1]{\begin{center}{ \sc #1} \end{center}}
\newcommand{\Address}[1]{\begin{center}{ \it #1} \end{center}}

\newcommand\pubblock{\rightline{\begin{tabular}{l} \pubnumber\\
         \pubdate  \end{tabular}}}
\newenvironment{Abstract}{\begin{quotation}  }{\end{quotation}}
\newenvironment{Presented}{\begin{quotation} \begin{center} 
             PRESENTED AT\end{center}\bigskip 
      \begin{center}\begin{large}}{\end{large}\end{center} \end{quotation}}
\newcommand{\Acknowledgements}{\bigskip  \bigskip \begin{center} \begin{large}
             \bf ACKNOWLEDGEMENTS \end{large}\end{center}}

\def\beq{\begin{equation}}
\def\eeq#1{\label{#1}\end{equation}}
\def\eeqn{\end{equation}}

\def\beqa{\begin{eqnarray}}
\def\eeqa#1{\label{#1}\end{eqnarray}}
\def\eeqan{\end{eqnarray}}

\let\bar=\overbar

\def\Dslash{\not{\hbox{\kern-4pt $D$}}}
\def\dslash{\not{\hbox{\kern-2pt $\del$}}}

\def\msb{{\bar{\ssstyle M \kern -1pt S}}}

\newcommand{\Sherpa}{S\scalebox{0.8}{HERPA}\xspace}

\newcommand{\Dire}{D\scalebox{0.8}{IRE}\xspace}  
\newcommand{\Csshower}{C\scalebox{0.8}{SSHOWER}\xspace}

\newcommand{\Pythia}{P\scalebox{0.8}{YTHIA}\xspace}

\newcommand{\Vincia}{V\scalebox{0.8}{INCIA}\xspace}

\newcommand{\Herwig}{H\protect\scalebox{0.8}{ERWIG}\xspace}

\newcommand{\Rivet}{R\scalebox{0.8}{IVET}\xspace}
\newcommand{\Professor}{P\scalebox{0.8}{ROFESSOR}\xspace}

\newcommand{\order}{\mathcal{O}}

\newcommand{\ttbar}{\ensuremath{t\bar{t}}}

\newcommand{\hl}{\vphantom{$\int_A^B$}}

\begin{document}
\begin{titlepage}
\pubblock

\vfill
\Title{Modelling and tuning in top quark physics}
\vfill
\Author{Marek Sch{\"o}nherr}
\Address{\institute}
\vfill
\begin{Abstract}
  In this proceedings I discuss the general strategy and impact of 
  tuning Monte-Carlo event generators for physics processes involving 
  top quarks. 
  Special emphasis is put on disinguishing the different usages of 
  event generators in the experiments and the subsequent implications 
  on the tuning process. 
  The current status of determining tune uncertainties is also discussed.
\end{Abstract}
\vfill
\begin{Presented}
$11^\mathrm{th}$ International Workshop on Top Quark Physics\\
Bad Neuenahr, Germany, September 16--21, 2018
\end{Presented}
\vfill
\end{titlepage}
\def\thefootnote{\fnsymbol{footnote}}
\setcounter{footnote}{0}

\section{Introduction}

Monte-Carlo event generators \cite{Buckley:2011ms,Sjostrand:2014zea,
  Bellm:2015jjp,Gleisberg:2008ta} are generally used in (at least) 
two fundamentally different ways:
\begin{itemize}
  \item[1)] They are used to calculate \emph{theory predictions}. 
            Here, parameters in the perturbative regime are 
            dictated by first principles or theory biases. 
            Parameters of models for non-perturbative physics are 
            determined universally in well-defined and limited sets 
            of observables (akin to PDF determinations).
  \item[2)] They are used for \emph{data modelling}. 
            To this end, all available parameters are tuned to best 
            reproduce the measured data of a specific process or in a 
            specific observable. 
            The resulting distributions lose all predictivity, but are 
            very useful to determine acceptances, efficiencies, systematic 
            correlations, etc.
\end{itemize}
Both cases are valid and needed, but must be clearly distinguished. 
This is especially relevant in the context of tuning the parameters 
of the models for non-perturbative physics employed by modern 
Monte-Carlo event generators.

This distinction becomes more relevant the smaller the target uncertainty 
of the prediction and the measurement become, and thus gains prominence 
in current and future cutting edge measurements in the top quark sector.
In this presentation, I highlight the standard paradigms of tuning 
Monte-Carlo event generators and will comment on how the uncertainty 
of these tunes can be assessed.

\section{Tuning strategies}

Monte-Carlo event generators are built by factorising collisions 
into different stages with different characteristic energy regimes, 
e.g.\ proton fragmentation, parton evolution, hard scattering, 
multiple interactions, hadronisation, hadron decays, etc, 
cf. Tab.\ \ref{tab:genstruc}. 
This factorisation also means that each stage is independent of the 
details of the other stages. 
For example, the hadronisation only depends on the colours, flavours 
and momenta of the parton ensemble at relatively small separations 
of $\order(\Lambda_\text{QCD})$. 
Each stage is then tuned individually as much as possible: 
Hadron decay parameters are first fitted to decay data from 
$b$-/$c$-factories. 
Then, the hadronisation parameters are tuned to $e^+e^-$ data 
at various energies ($b$-factories, SLD, LEP), before the parameters 
of the multiple interaction model, beam remnant parametrisation, etc.\ 
are tuned to hadron collider data. 
In all stages \Rivet \cite{Buckley:2010ar} and \Professor \cite{Buckley:2009bj} 
are the commonly used tools.

\begin{table}[t!]
  \centering
  \begin{tabular}{r||l|l|l}
    & \textbf{\Herwig7} & \textbf{\Pythia8} & \textbf{\Sherpa} \hl\\
    \hline\hline
    \multirow{3}{*}{PS} 
    & $\tilde{q}$-Shower & Default Parton Shower & \Csshower \hl\\
    & Dipole Shower & \Dire & \Dire \hl\\
    & & \Vincia & \hl\\
    \hline
    \multirow{3}{*}{MI} 
    & soft gluon model  & sophisticated & old \Pythia-style \hl\\
    & \& hard scat.\ model & interleaved model & non-interleaved \hl\\
    & (\textsc{Jimmy}-based) & & \\
    \hline
    \multirow{3}{*}{Had} 
    & Cluster & Lund String & Mod.\ Cluster \hl\\
    & Interface to & & Interface to \hl\\
    & Lund String & & Lund String \hl\\
    \hline
    MB 
    & MinBias & MinBias & -- \hl
  \end{tabular}
  \caption{
    List of parton showers and non-perturbative models employed by 
    multi-purpose event generators.
    \label{tab:genstruc}
  }
\end{table}

Tuning the non-perturbative models of an event generator generally 
bases on the minimisation of the following definition of $\chi^2$ 
dependent on a set of parameters $\vec{\mathbf{x}}$
\begin{equation}
  \chi^2(\vec{\mathbf{x}})
  =\frac{1}{N}\sum\limits_{i\in\mathcal{O}}w_i\,
    \frac{\left(\text{MC}_i(\vec{\mathbf{x}})-\text{Data}_i\right)^2}
	{\sigma_{i,\text{Data}}^2}\;,
\end{equation}
wherein $N$ is the number of observables the index $i$ runs over, 
$\text{MC}_i(\vec{\mathbf{x}})$ is the Monte Carlo prediction given 
the parameter set $\vec{\mathbf{x}}$, $\text{Data}_i$ is the 
measured data and $\sigma_{i,\text{Data}}$ is its uncertainty
\cite{Buckley:2009bj,Skands:2014pea}.

\begin{figure}[t!]
  \centering
  \includegraphics[width=0.47\textwidth]{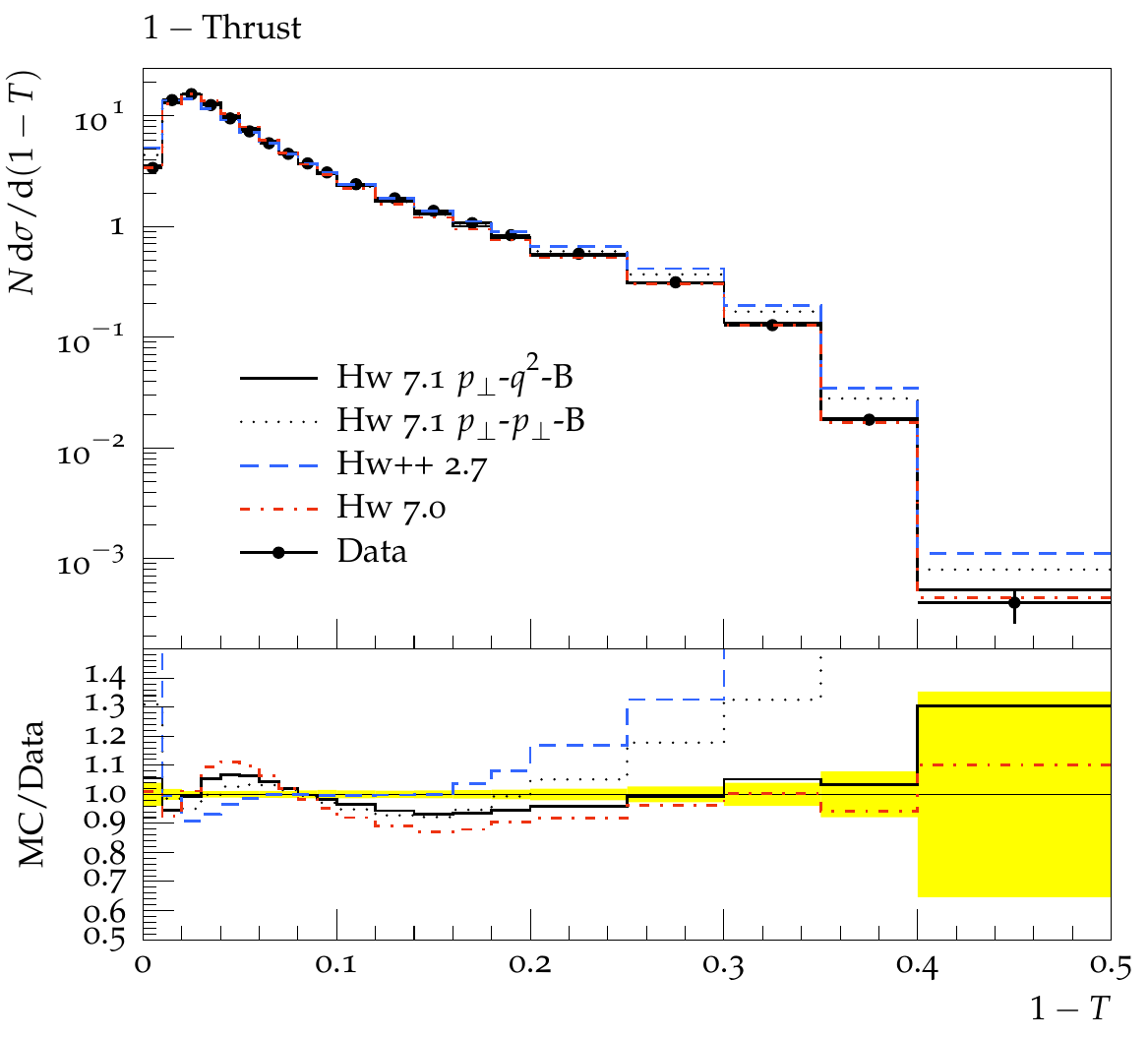}
  \hfill
  \includegraphics[width=0.47\textwidth]{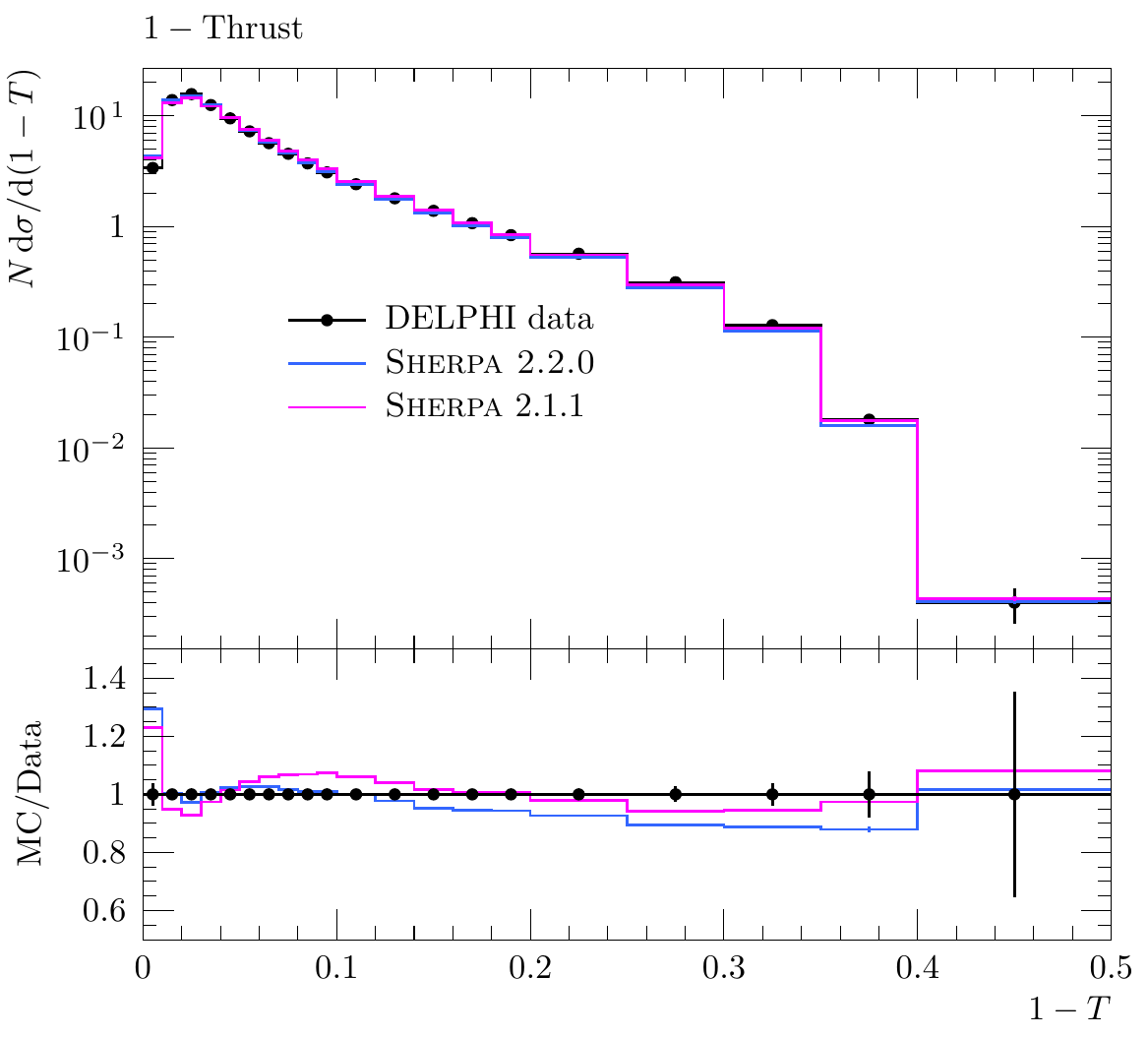}
  \\
  \caption{
    Example of the quality of tuning of different releases of 
    the \protect\Herwig (left) and \protect\Sherpa (right) 
    generators compared to data taken by the DELPHI coll.\ 
    \cite{Abreu:1996na}.
    \label{fig:tuning-herwig-sherpa}
  }
\end{figure}

\begin{figure}[t!]
  \centering
  \includegraphics[width=0.47\textwidth]{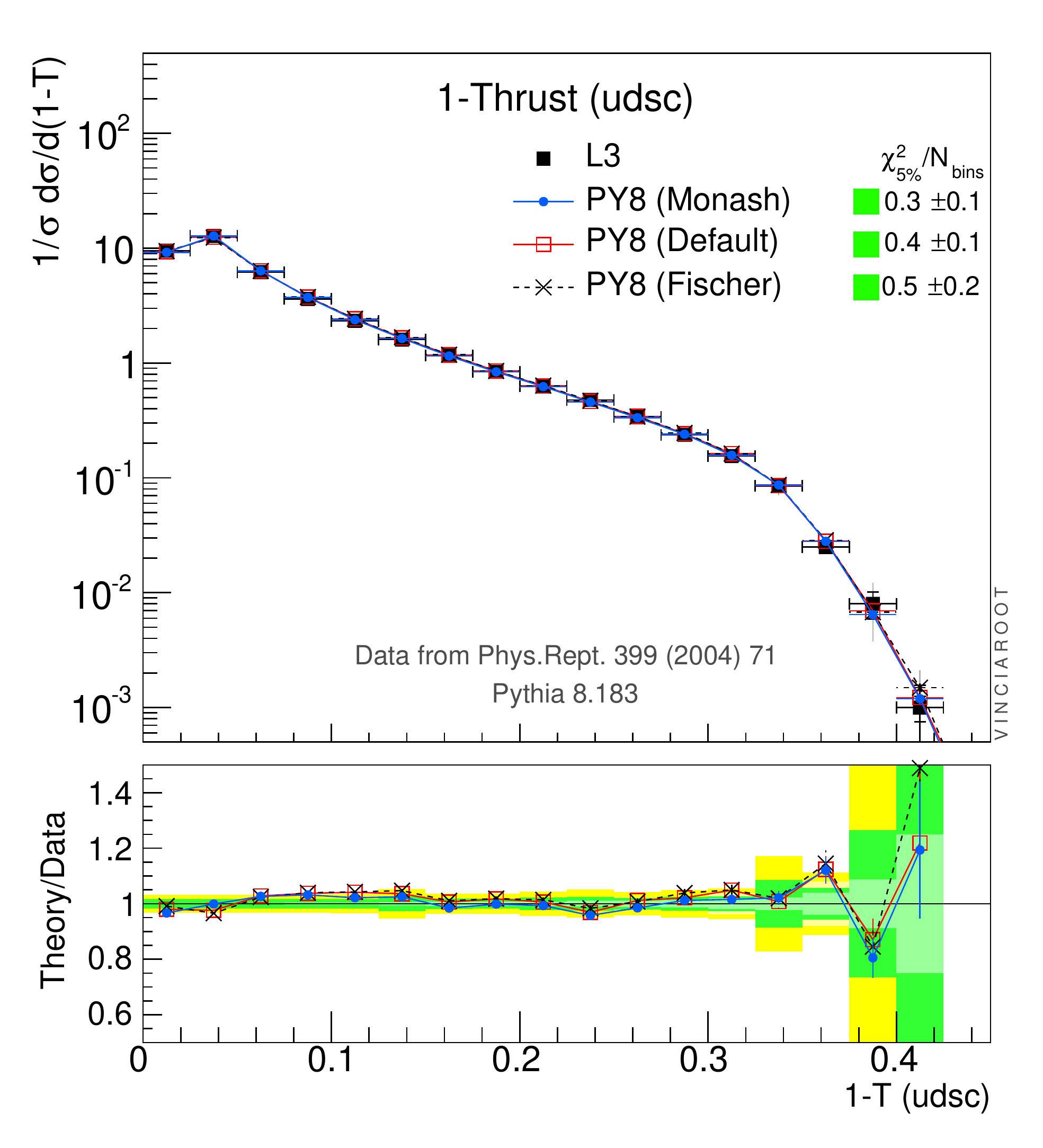}
  \hfill
  \includegraphics[width=0.47\textwidth]{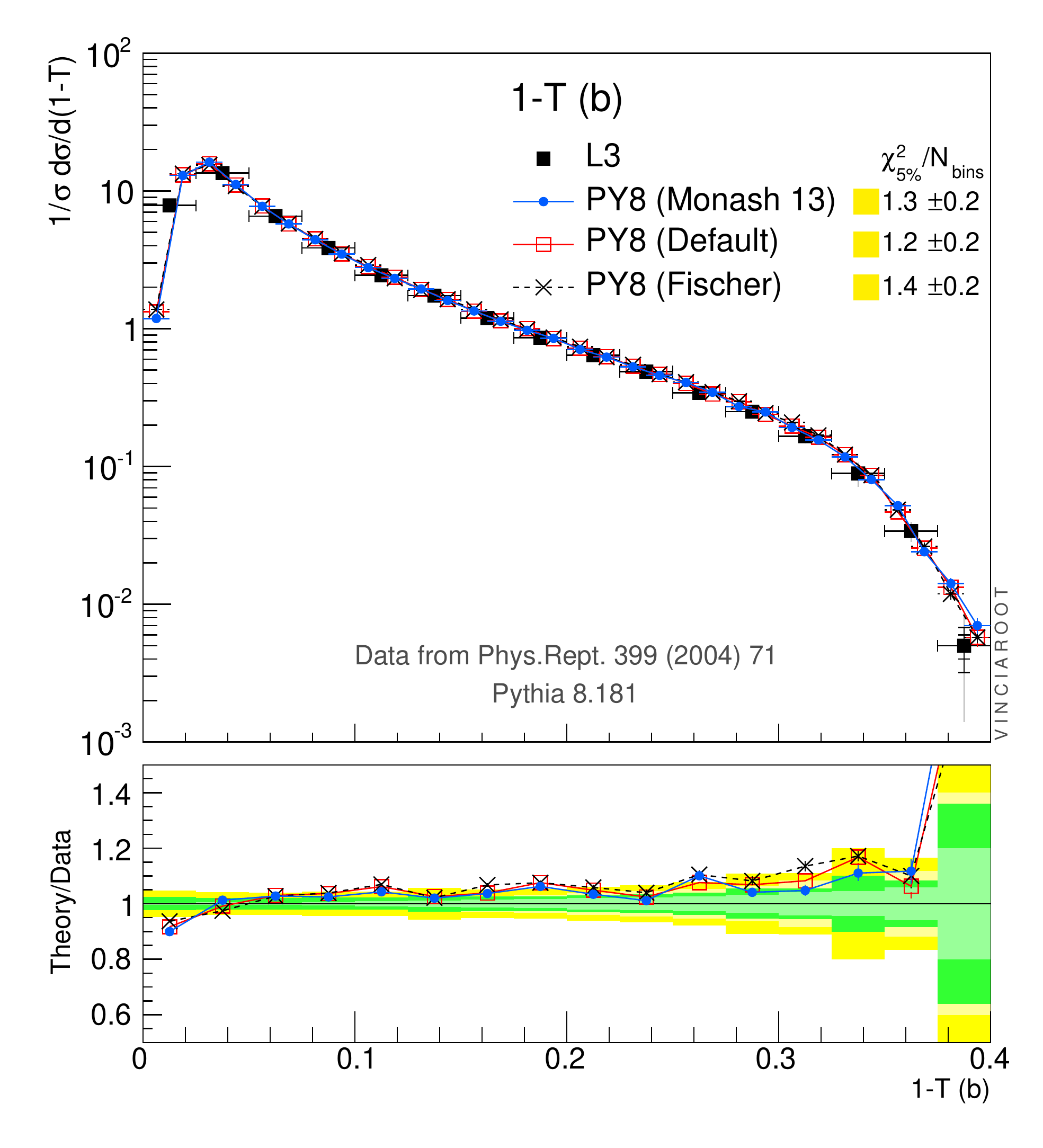}
  \\
  \caption{
    Example of the quality of tuning in \protect\Pythia8 
    for light (left) and $b$ jets (right) compared 
    to data taken by the L3 collaboration \cite{Achard:2004sv}. 
    Figures taken from \cite{Skands:2014pea}.
  }
\end{figure}

As the standard setups used to tune the non-perturbative 
models only include limited perturbative information, 
caution must be applied when including observables that receive 
sizeable contributions from multijet final state. 
They, thus, can only be included if the physics that should be 
modelled perturbatively is included properly before engaging 
the non-perturbative event phases. \cite{ATL-PHYS-PUB-2013-017}

\begin{figure}[t!]
  \includegraphics[width=0.47\textwidth]{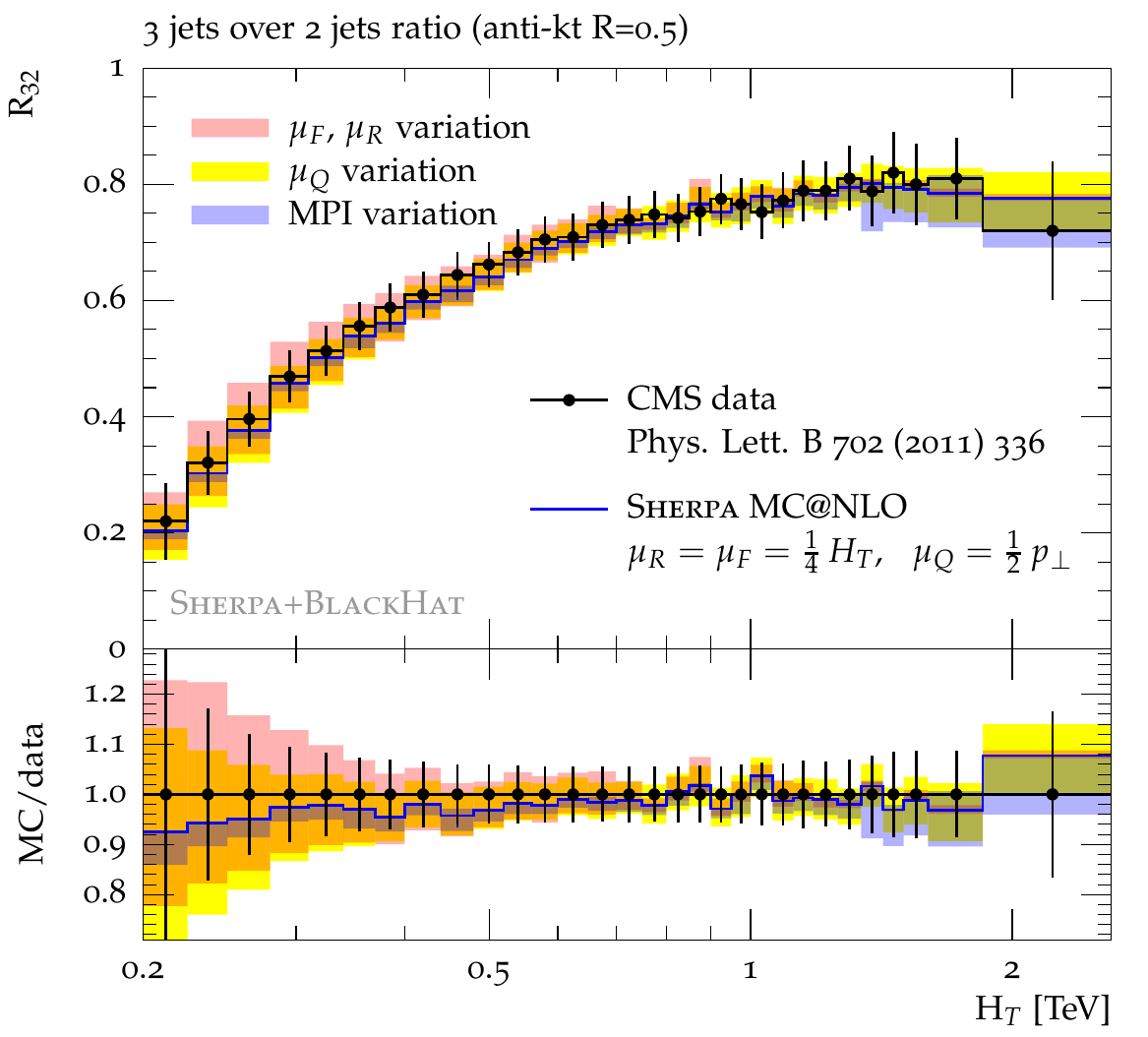}
  \hfill
  \includegraphics[width=0.47\textwidth]{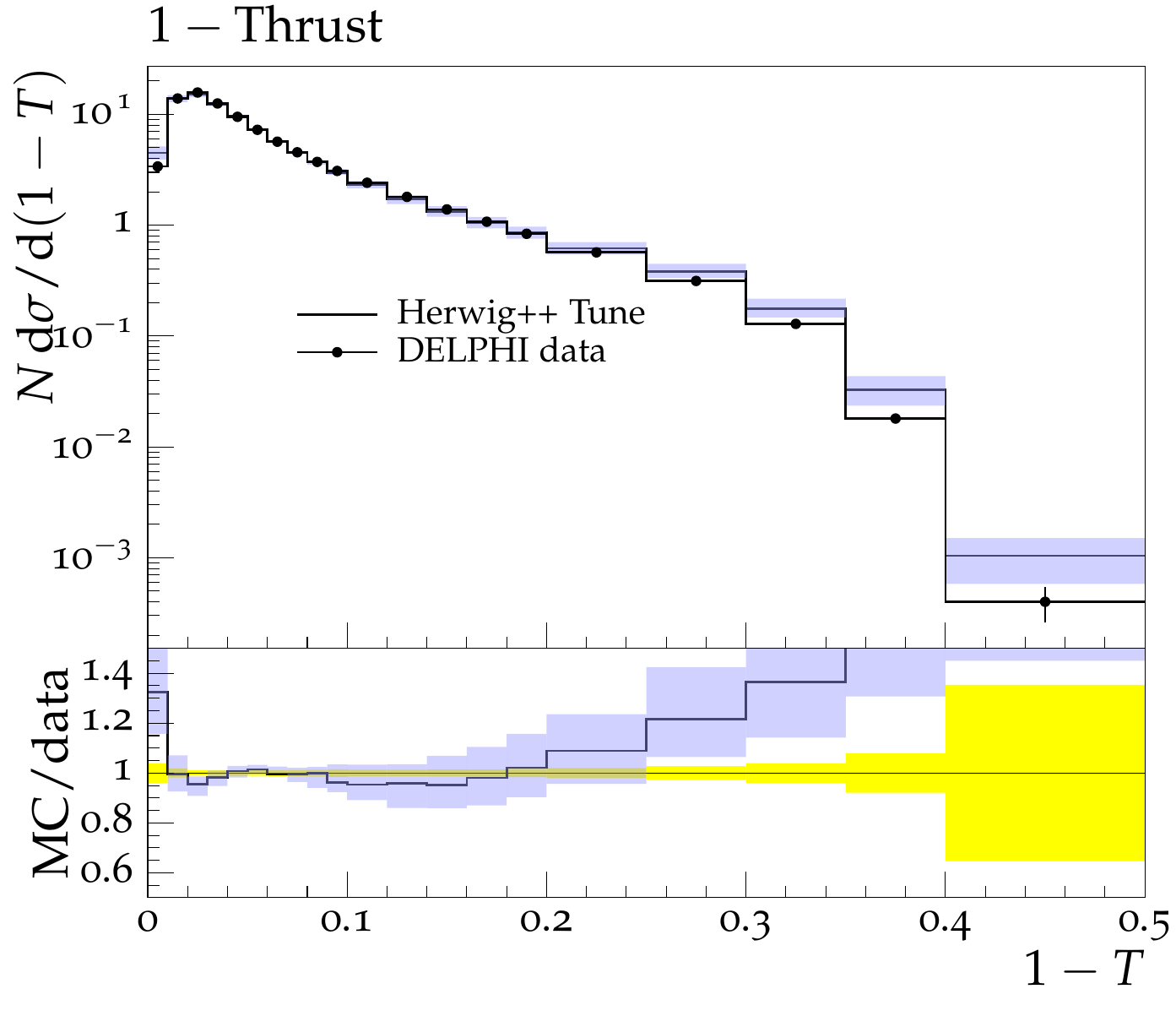}
  \caption{
    Example of tune uncertainties defined through ``reasonable'' 
    variations of selected model parameters (left, for the $R_{32}$ 
    at the LHC) and systematically defined Eigentunes (right, for $1-T$ 
    at LEP).
    Left figure taken from \cite{Hoche:2012wh}.
    \label{fig:tune-uncs-1}
  }
\end{figure}

Traditionally, expert-defined variations of certain key parameters 
within ``reasonable'' ranges are used as simple stand-ins to 
gauge the uncertainty of the given tune, i.e.\ minimum point 
$\vec{\mathbf{x}}$ of the $\chi^2(\vec{\mathbf{x}})$ distribution. 
Therein, the ``reasonable'' range is usually defined such that 
the resulting uncertainty is comparable to the data uncertainty in 
the measurement \cite{AlcarazMaestre:2012vp}. 
This can be improved by assuming that the test statistics were 
distributed according to a second order polynomial around the 
above minimum or optimal tune. 
A variation around this minimum to some $\Delta\chi^2$ then 
forms an ellipsoid with $2\,\text{dim}(\vec{\mathbf{x}})$ principal 
vectors forming the eigenvectors of the covariance matrix 
computed around the minimum. 
Variations with the customary $\Delta\chi^2=1$, defining 
a $1\,\sigma$ variation under the above assumption, however, 
lead to empirically too small variations. 
Thus, a more useful value of 
$\Delta\chi^2=\tfrac{1}{2}\,\text{dim}(\vec{\mathbf{x}})$ 
is adopted in practise \cite{Buckley:2018wdv}. 
Future approaches, relaxing the above assumptions of the 
behaviour of the models around the minimum, are currently 
being developed and will lead to more agnostic definitions 
of tuning uncertainties, necessitating less arbitrary definitions 
of ``reasonableness''.

\section{Influence on top quark observables}

\begin{figure}
  \includegraphics[width=0.47\textwidth]{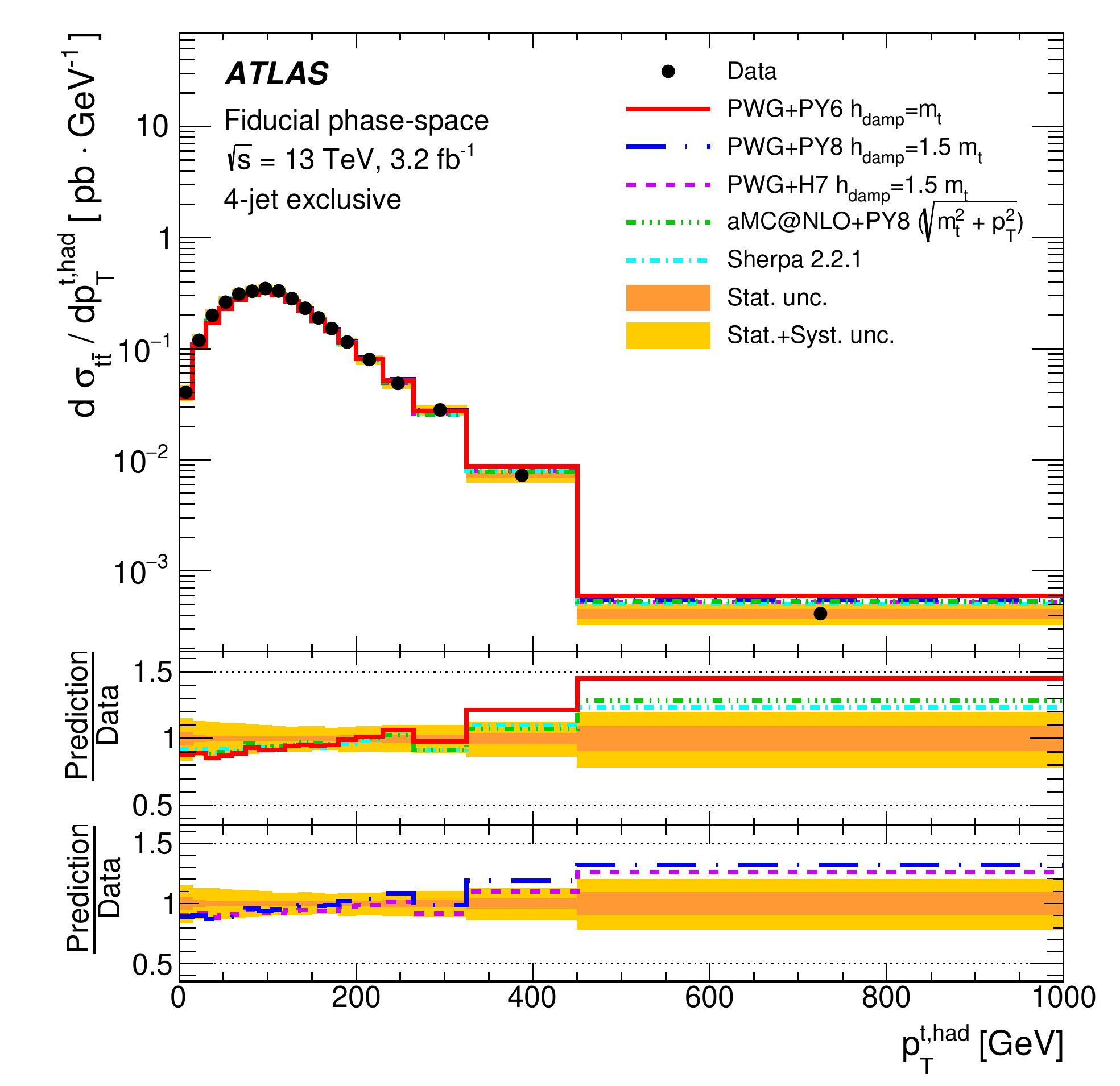}
  \hfill
  \includegraphics[width=0.47\textwidth]{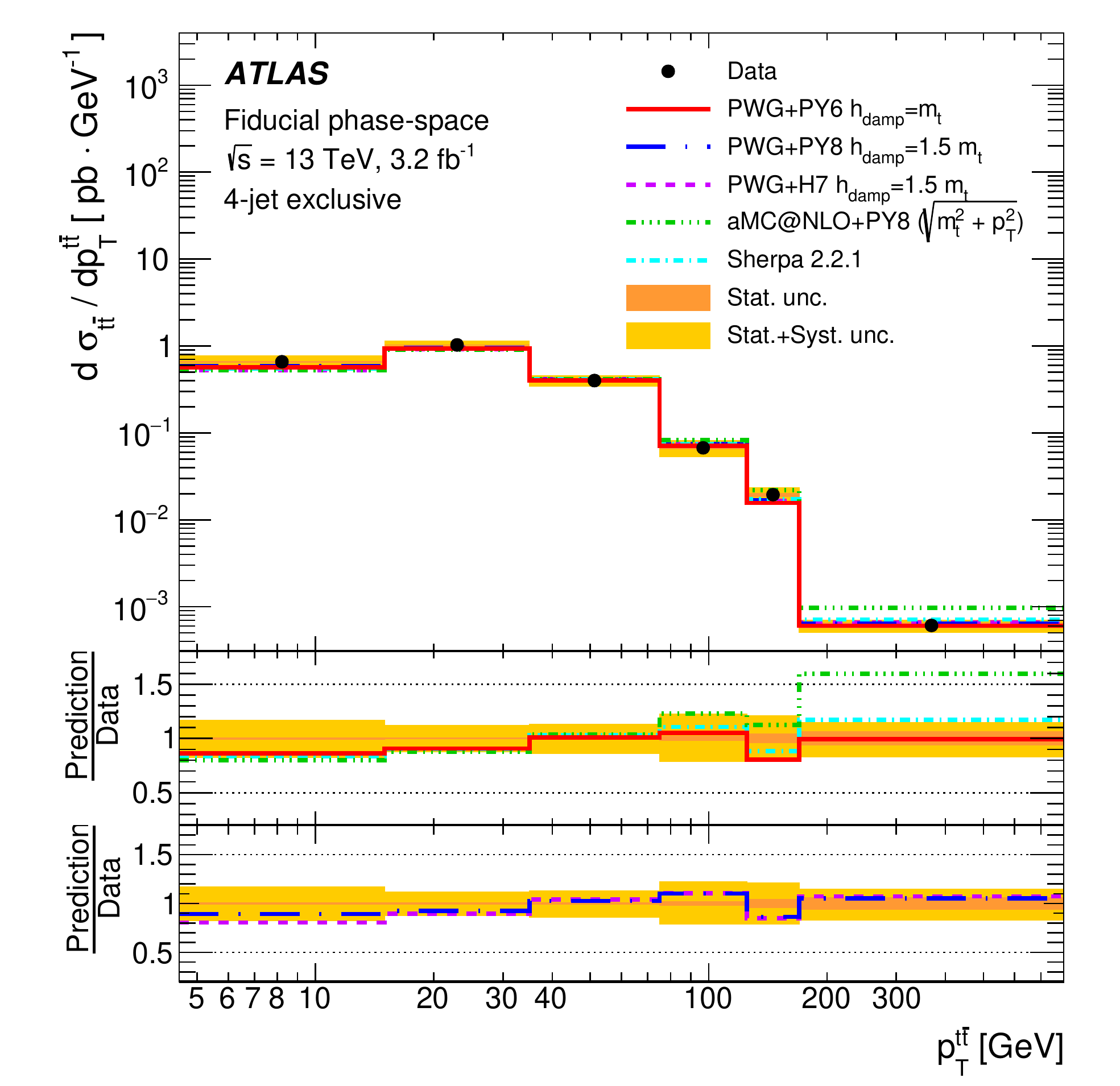}
  \caption{
    General $t\bar{t}$ observables show only very little dependence 
    on the details of the non-perturbative models and are 
    well described by current tunes.
    Figures taken from \cite{Aaboud:2018uzf}.
    \label{fig:atlas-ttbar}
  }
\end{figure}

In contrary to most data used in tuning Monte-Carlo event 
generators, top quark processes are unique in that they 
always involve $b$-quarks and almost always involve gluons. 
Both are not as constrained as their light quark counter parts. 

Most observables in standard $t\bar{t}$ measurements show only 
very little dependence on the details of the non-perturbative 
modelling. 
Subsequently, the uncertainty of the current global tunes of 
all modern event generators hardly impact these measurements. 
Nonetheless, dedicated examples can be found where they impact. 
To reduce the tuning uncertainty on this class of observables, 
one may be tempted to include top quark observables in the tunes. 
This would, however, exclude exactly these observables from 
collection of observables unbiased theoretical predictions 
can be made for. 
A different approach would be to find suitable proxies to better 
constrain the gluon and bottom related parameters of the 
non-perturbative models. 
Such measurements are already on their way ($\Delta R(B,B)$ at the LHC 
\cite{Khachatryan:2011wq,Aaboud:2018uiu} for $t\bar{t}b\bar{b}$ observables, 
gluon jet data at LEP \cite{Fischer:2014bja,Fischer:2015pqa} and LHC 
for gluon jet fragmentation, etc.), but will need to be more routinely 
used in global tuning efforts.

Not all cases of mismodelling are related to imperfect tuning 
or missing aspects in the non-perturbative models though. 
The transverse momentum distribution of the top quark in 
\ttbar production can be resolved by including electroweak 
corrections, at least in the large transverse momentum region, 
as shown in \cite{Gutschow:2018tuk}.

\section{Conclusions}

It is important to clearly distinguish (at least) two types of 
Monte-Carlo event generator usage: calculating theory predictions 
and providing a controllable and fast tool for data reproduction. 
While both use cases are perfectly valid in their own right, they 
mandate different treatments of the tuning of the parameters of 
their models of non-perturbative physics. 
If the Monte-Carlo event generator is used to best reproduce 
already measured data, e.g.\ for statistical analysis or the 
evaluation of internal correlations, all parameters of the 
non-perturbative physics models (and some of the perturbative 
calculation) may be adjusted to best fit the data of that 
measurement. 
Of course, by definition, the resulting output of the Monte-Carlo 
event generator is no theoretical prediction in this case but 
fulfils its role as multidimensional fit function. 
If, on the other hand, a theoretical prediction is sought, 
the parameters of the non-perturbative models must be tuned 
to a well-defined and limited global set of observables,
while those of the perturbative calculation should be set 
to a theoretically defined value. 
The resulting tunes aim to describe all available data to the 
best ability of the employed models are can be used for theory 
predictions for all observables except the input observables.

Currently, no tune provided by the authors of the used Monte-Carlo 
event generators makes use of any direct top quark data. 
Thus, these tunes can be employed for theory predictions for 
top quark observables. 
Of course, when observables receive a non-negligible non-perturbative 
contribution, mismodelling may happen as the employed models are 
phenomenological in nature and do not directly derive from first 
principles and can, thus, not fully capture the underlying dynamics. 
Conversely, mismodelling does not necessarily originate in suboptimal 
tuning or incomplete models and the impact of improved perturbative 
inputs should not be underestimated.

Uncertainties on existing tunes can currently only be assessed using 
some level of arbitrary ``reasonableness'' criterion, either through 
a hand-picked set of representative variations or a set of Eigentunes 
where the magnitude of the $1\,\sigma$ ellipsoid is set manually. 
New methods along the lines of using data replica to define alternative 
tunes are currently being explored.

\Acknowledgements
MS acknowledges the support of the Royal Society through the award 
of a University Research Fellowship.

\end{document}